\documentclass{aa}
\usepackage{graphics}

\newcommand{\ale}{\alpha_{\rm e} }
\newcommand{\alr}{\alpha_{\rm r} }
\newcommand{\alinj}{\alpha_{\rm inj} }

\begin{document}
\thesaurus{08.16.6; 09.09.1 Crab Nebula; 09.19.2; 13.18.3}
\title{Radio spectrum of the Crab nebula as an evidence
for fast initial spin of its pulsar}
\author{ A. M. Atoyan \inst{1,2}}
\institute{Max Planck Institut f\"ur Kernphysik, Saupfercheckweg 1,
           D--69117 Heidelberg, Germany; armen.atoyan@mpi-hd.mpg.de 
\and Yerevan Physics Institute, 375036 Yerevan, Armenia}

\date{Received; accepted}
\titlerunning{Radio spectrum of the Crab nebula}
\maketitle

\begin{abstract}
 The origin of relativistic electrons in the Crab nebula 
which are producing the broad-band flat radio
spectrum of this prototype plerion has proved difficult to understand.
Here I show that these electrons can be naturally explained as a relic 
population of the pulsar wind electrons 
 that  have lost most of their energy in the expanding nebula
because of intensive radiative and adiabatic
cooling in the past. The observed radio spectrum 
suggests that the initial slowing-down time of the pulsar 
was $\tau_{\rm sd}\leq 30\,\rm yr$, which implies   
that it has been born with a spin period $P_0\sim 3-5\,\rm ms$, several  
times shorter  
than presently believed. Consistency of these results with the current data and 
the historical records is discussed.

\end{abstract}

\section{Introduction}

Flat radio spectra, with power-law indices $0\leq \alpha_{\rm r} \leq 0.3$
(for $S_{\nu}\propto \nu^{-\alpha}$\,) represent
one of the key features of the plerions (Weiler \& Panagia 1978),   
that distinguish these filled-center supernova remnants (SNRs)   
from a typical shell-type
SNR with the mean  $\alr \sim 0.5$ (e.g. Green, 1991).
Meanwhile, the origin of the electrons producing these unusual radio 
spectra  is not yet understood (e.g. Green and Scheuer 1992, 
Woltjer et al. 1997). 

Radio electrons appear particularly enigmatic 
for the Crab Nebula (see Kennel \& Coroniti 1984 -- hereafter KC84)
where the synchrotron spectrum    
extends with $\alr \simeq 0.3$ (Baars et al. 1977)  
from $ 10^7\,\rm Hz$ to 
 $ 10^{13}\,\rm Hz$, and then it steepens to $\alpha_{\rm opt}\sim 0.8$ 
in the IR/optical band (Marsden et al. 1984) This requires
an energy distribution of the electrons 
$N(\gamma)\propto \gamma^{-\ale}$ with
$\ale=1+2\,\alr\simeq 1.6$ extending over 3 decades of  energy  
 (Lorentz factor) $\gamma $ 
from  $ \sim \! 300 $ to $3\times 10^5$, which then steepens 
to $\ale\simeq 2.6$.
For the mean magnetic field $B_\ast \simeq 0.3\,\rm mG$ and the known 
age $t_{\ast}$ of the Crab Nebula, born in 1054, 
this steepening  seems to agree with the radiative break $\Delta \alpha =0.5$ 
(Kardashev 1962) both in the magnitude
and position of the break.
 One might thus suppose that all relativistic electrons in the nebula,
including the ones with $\gamma$ below the 
 radiative break energy $\gamma_{\ast \rm br} \sim 3\times 10^5$ (at present), 
called "radio electrons", represent
a single population of particles currently accelerated 
at the pulsar wind termination shock (Rees \& Gunn 1974 -- RG74)
that would produce an injection spectrum 
$Q(\gamma)$ with 
 $\alinj =1.6$ at $\gamma \leq 3\times 10^6$, and 
$\alinj\simeq 2.3$ at higher energies (in order to explain the 
X-ray fluxes with $\alpha_{\rm x}\simeq 1.1-1.2$). 
However, the mean electron energy  in such a spectrum
$\overline{\gamma}\leq 10^4$, which is much less than the energy 
of the wind electrons upstream of the shock, $\gamma_{\ast \rm w}\sim 10^6$  
(e.g. Kundt \& Krotscheck 1980 -- KK80, KC84, Arons 1996 -- Ar96), whereas the 
energy and particle number flux conservation laws across the shock 
 require that $\overline{\gamma}\sim \gamma_{\ast \rm w}$. 
An idea of current acceleration of radio electrons in the main nebula 
beyond the wind termination shock 
appears also problematic, because radio observations do not show 
any significant variation
of the radio spectral index, implying an absence of  effective electron   
acceleration sites there (Kronberg \& Bietenholz 1992, 
Bietenholz et al. 1997).

The remaining alternative is to assume that  radio electrons in the Crab 
nebula are relics of the history of its evolution (Shklovskii 1977, KK80, 
KC84). The model proposed below reconsiders the evolution
of the electrons injected into the expanding nebula at the pulsar wind shock.

\section{Results}

Previous studies of  
expanding plerions (Pacini \& Salvati 1973 -- PS73; 
Reynolds \& Chevalier 1984 -- RC84) have assumed an 
injection spectrum  $Q(\gamma,t) \propto \gamma^{-\alinj}$
with $\alinj=1+2\alr$ starting from $\gamma\geq 1$. Thus,    
the origin of such a flat spectrum was not considered. 
In agreement with theoretical predictions 
(KK80, KC84, Ar96, Atoyan \& Aharonian 1996), here I assume an injection spectrum 
with a profound deficit of electrons below 
some $\gamma_{\rm c}$, for 
example in the form  $Q(\gamma,t) \propto x^2(1+x)^{-2 -\alpha_{\rm inj}}
\,\exp (-\gamma/\gamma_{\rm max})\; $ 
where  $\,x= \gamma/\gamma_{\rm c}$. For  $\alinj\approx 2.3$, 
 the mean energy of electrons in this spectrum  
$\overline{\gamma} \simeq (8-10)\gamma_{\rm c}$. This defines    
 $\gamma_{\rm w}$ of the pulsar driven wind of 
electromagnetic fields and particles which powers the plerion (RG74).
The electrons are injected into the nebula with a power 
$L_{\rm e}=\eta_{\rm e}L_{\rm sd}$ where  
$L_{\rm sd}= L_{0} /(1+t/\tau_{\rm sd})^k $ is the spin-down luminosity, 
and $k=(n+1)/(n-1)$ in terms of the pulsar braking index $n$.  
A significant, if not the dominant 
($\eta_{\rm e}\rightarrow 1$, KC84),  fraction of $ L_{\rm sd}$ is injected in
the electrons, so  
$\gamma_{\rm w}(t)\propto L_{\rm sd}/\dot{N}$ where
$\dot{N}(t)$ is the  production rate of $e^{+}-e^{-}$ pairs by the pulsar.
Very generally, $\dot{N}$ can depend mainly on the magnetic field $B_{\rm n}$ of 
the neutron star and its spin  frequency 
$\Omega(t)=\Omega_0 (1+t/\tau_{\rm sd})^{-1/(n-1)}$.
Because the basic time dependent parameter is 
$\Omega$,
I approximate $\gamma_{\rm w}\propto \Omega^{p}(t)$ with a 
parameter $p$.  This results in 
$\gamma_{\rm w} \propto (1+t/\tau_{\rm sd})^{-s}$ with\footnote{Note that 
this relation predicts $\gamma_{\rm w}\simeq \rm const$ for times
 $t\ll   \tau_{\rm sd}$. 
This seems to be  correct also generally, allowing a reasonable variation
(e.g. a gradual decline) of $B_{\rm n}(t)$, although this would 
somewhat change the relation $s=p/(n-1)$.} a model parameter $s=p/(n-1)$.   

Because of intensive synchrotron losses in the past, electrons were brought 
to energies much smaller than $\gamma_{\ast \rm br}$ at present.  
For the energy losses  
$P_{\rm s}\equiv -(\partial \gamma /\partial t)_{\rm s} = b\gamma^2$ where
$b\approx 0.04 (B/1\,\rm G)^{2} \;\rm yr^{-1}$, and 
magnetic field in the expanding nebula declining in time as 
$B=B_\ast (t_\ast/t)^m$, typically with $m \sim$\,1.3-2 (RC84), 
the equation $\gamma/P_{\rm s}=t$ gives the radiative brake energy   
\begin{equation}
\gamma_{\rm br}(t)=3\times 10^{5} (B_{\ast}/0.3 \,\rm mG)^{-2}
\,(t/t_\ast)^{2m -1} \; .
\end{equation} 
All electrons injected into the nebula at times $t$ with energies higher than
$\gamma_{\rm br}(t)$ are rapidly cooled down to 
this energy. Electrons of smaller energies are affected only by adiabatic
energy losses which do not change a power-law shape of $N(\gamma)$ once 
established.

The origin of the flat distribution of radio electrons in plerions 
is readily understood if we neglect, for a moment,
the adiabatic energy losses. Then
Eq.(1) predicts that all electrons that enter the
nebula between times $t$ and $t+\Delta t$ with energies above 
$\gamma_{\rm br }(t)$ will be soon 
found in the energy region between $\gamma\approx \gamma_{\rm br }$ and
$\gamma+\Delta\gamma$, where $\Delta\gamma\propto t^{2m-2}\Delta t$. 
Since injection occurs mostly at high energies,
a $\delta$-functional approximation 
$Q(\gamma,t)= \dot{N}\delta(\gamma-\gamma_{\rm w})$, where 
$\dot{N} =\dot{N}_0 (1+t/\tau_{\rm sd})^{s-k}$,
can be used to find the number of electrons $\Delta N =\dot{N} \Delta t$
accumulated in the interval $\Delta \gamma$. Eq.(1) connects
the injection time of electrons with their energy after radiative cooling as 
$t\propto \gamma^{1/(2m-1)}$. Then the distribution 
$\Delta N/\Delta \gamma \rightarrow N(\gamma) $ of electrons that have been 
injected at times $t\ll \tau_{\rm sd}$ acquires a power-law slope 
$\propto \gamma^{-\alpha_{\rm e}}$ 
with $\ale=1-1/(2m-1)$. At time $t\sim \tau_{\rm sd}$ the injection rate
$\dot{N}(t)$ starts to decline, resulting in  an increase of the slope  
to $\ale \sim 1$. This can explain the spectra of plerions like 3C 58 or 
G21.5-0.9 where $\alr\sim 0$, and suggests that such plerions are 
powered by pulsars with $\tau_{\rm sd}$ 
not much less than their age. However, radio spectra with 
$\alr \simeq (0.2-0.3)$ are produced by the electrons injected at 
times  $t\gg \tau_{\rm sd}$, when $\dot{N}(t)$ rapidly declines, and  
$\ale \rightarrow 1 +(k-s-1)/(2m-1)$. This  
slope extends up to the current radiative break energy 
$\gamma_{\ast \rm br}$, if 
the energy  of electrons in the wind presently   
$\gamma_{\ast \rm w} > \gamma_{\ast \rm br}$, as in Crab nebula,  
and to a smaller energy otherwise.

 This qualitative analysis demonstrates the role of  the radiative losses  
for the formation of flat spectra of radio electrons in plerions. The results  
do not significantly change after  we 
take into account also the adiabatic energy losses of electrons,
$P_{\rm adb}=a\,v\gamma/R$. In the approximation of a homogeneous spherical
source of a radius $R$ which expands with a velocity $v$, the parameter $a=1$
(Kardashev 1962). For an arbitrary $a$, the energy distribution of electrons  reads
(Atoyan \& Aharonian 1999):  
\begin{equation}
N(\gamma,t)=\int_{0}^{t} \,
\left( \frac{R_1}{R} \right)^{a} \frac{\Gamma_{t_1}^2}{ \gamma^2}
\, Q(\Gamma_{t_1},t_1) \, {\rm d} t_1\; ,
\end{equation}
where $R_1\equiv R(t_1)$, and 
$\Gamma_{t_1} \equiv \Gamma(\gamma, t, t_1)$ is the particle energy 
at an instant $t_1\leq t$: 
\begin{equation}
\Gamma(\gamma,t,t_1) =
\gamma \left( \frac{R}{R_1}\right)^{a}
\left[ {1-\,\gamma \,
\int_{t_1}^{t} b(t_2)
\left(\frac{R}{R_2}\right)^{a} \, {\rm d}t_2 }\right]^{-1} \cdot
\end{equation}
Expansion of the Crab nebula is approximated as 
$v=v_{\ast} (t/t_\ast)^\delta$ with acceleration
parameter $\delta \approx 0.1$, as observed (Trimble 1968).
In the region 
$\gamma_{\rm sd} \ll \gamma \leq \gamma_{\ast \rm br}$, where 
$\gamma_{\rm sd}\approx \gamma_{\ast \rm br} \, 
(\tau_{\rm sd}/t_{\ast})^{2m -1+a+a\delta}$ is the contemporary energy of the 
electrons injected at time $t= \tau_{\rm sd}$, the    
 $\delta$-functional approximation for $Q(\gamma)$  
results in $N(\gamma,t_\ast)\propto \gamma^{-\ale}$ with
\begin{equation}
\ale = 1 +(k-s-1)/(2m -1 +a+a\delta)\; ,
\end{equation}
 which confirms the result
derived qualitatively above for $a=0$. 
The spectral index $\alr=(\ale-1)/2$ then implies    
\begin{equation}
s=k-1 - 2\alr \, (2 m  -1 +a + a\delta)\; . 
\end{equation}

The energy distribution of radio electrons in the Crab nebula
should extend with the same index $\ale$ over 3 decades of energy below 
$\gamma_{\ast \rm br}$. This requires
that $10^{-3}\gamma_{\ast \rm br}\gg \gamma_{\rm sd}$, therefore
the radio spectrum of the Crab nebula would suggest    
$\tau_{\rm sd}\ll 10^{-3/(2m-1+a+a \delta)} t_{\ast} \simeq 100\,\rm yr$ (!).

\begin{figure}[htbp]
\resizebox{8.5cm}{!}{\includegraphics{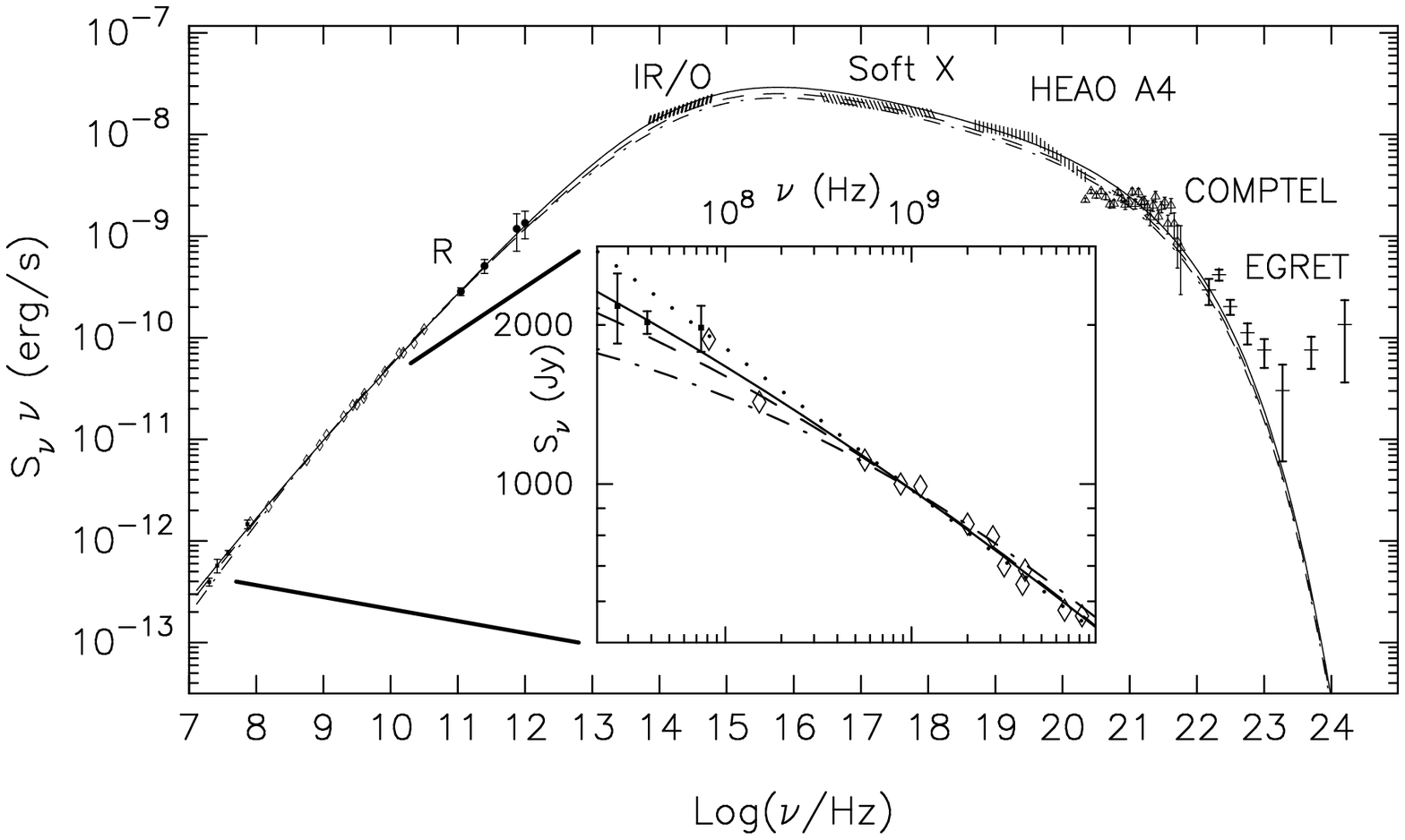}}
\vspace{3mm}

\hspace{5mm}\resizebox{7.5cm}{!}{\includegraphics{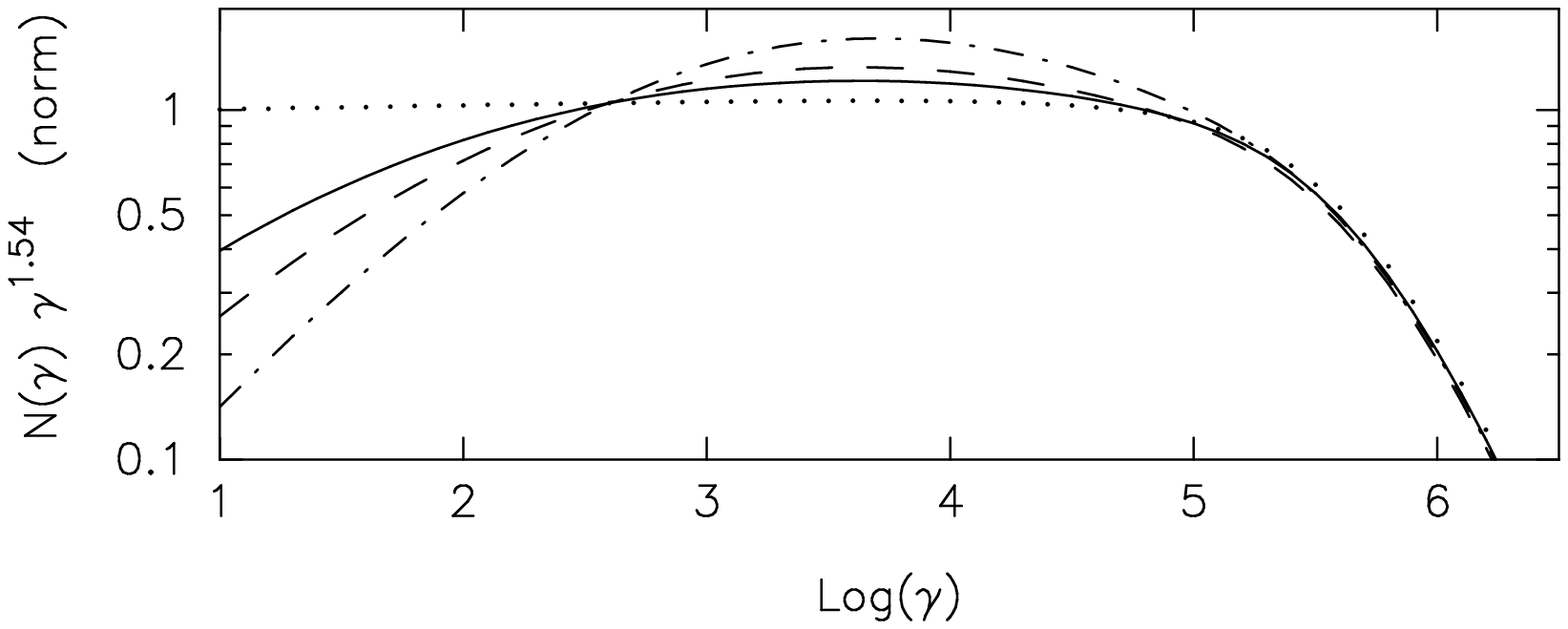}}
\caption{The radiation fluxes (top) and the distribution of electrons 
(bottom) 
calculated for different  $\tau_{\rm sd}$. The
parameter $s$ has been adjusted to give the best possible fit to radio data: 
$s=-0.7$ for $\tau_{\rm sd}=20\,\rm yr$ (dashed); 
$s=-1.1$ for $\tau_{\rm sd}=40\,\rm yr$ (solid);
$s=-2$ for $\tau_{\rm sd}=100\,\rm yr$ (dot-dashed).
The dotted lines are calculated in the formal case of 
$\tau_{\rm sd} =1\,\rm yr$ assuming $s=-0.3$, in agreement with  Eq.(5). 
Other model parameters are fixed at $m=1.4$, $a=1$, $k=2.3$, $\delta =0.1$.
The fit at the optical and higher frequencies is reached assuming  
$\alinj=2.3$, $\gamma_{\rm max}=2.2\times 10^9$, and 
$\gamma_{\rm c}=1.9\times10^5$ (the principal parameter actually is the mean   
$\overline{\gamma}\approx 1.6\times 10^6$).
Compilation of data is from Atoyan \& Aharonian (1996) and references therein. 
The size of the diamonds in the inset corresponds
to a typical accuracy $\pm 6\,\%$  of the fluxes reported by Baars et al (1977).
Note that the negative values of $s$  correspond to  
a gradual {\it increase} of $\gamma_{\rm w}$ in time. 
Qualitatively, this can be explained as a result of very high efficiency
of the cascade (Sturrock 1971) of $e^{+}-e^{-}$ 
pairs and $\gamma$-rays in the pulsar 
magnetosphere. In the past, when the pulsar rotated faster, 
this cascade was developing in a more compact ($\propto c/\Omega$ ) and 
tense magnetosphere, where electrons were effectively produced in the 
thus much higher magnetic fields. 
Fast increase of $\dot{N}$ would then result in a smaller $\gamma_{\rm w}$. 
 }
\end{figure}

Numerical calculations confirm this very unexpected and far reaching  
conclusion. It  may seem 
at the first glance to Fig.~1 that all curves explain 
equally well the measured synchrotron fluxes 
from radio to optical, and even further to X-ray and
$\gamma$-ray regions (where the fluxes can be sensitive to 
spatially inhomogeneous distribution of the nebular magnetic field). 
However, the radio fluxes of the Crab nebula are known, and must be explained,
with high accuracy $\leq 10\%$.
A closer look to the fluxes $S_{\nu}$ shown on the inset in Fig.1
reveals that an explanation of the data at low frequencies 
is acceptable only if  
$\tau_{\rm sd} \sim 30\,\rm yr$ or smaller.  
This implies that the spin frequency $\Omega$ of the Crab pulsar
 initially was $(1+t_\ast/\tau_{\rm sd})^{1/(n-1)}\approx
10$ times higher than presently.
For the spin period $P=2\pi/\Omega$ this corresponds to
the initial $P_0 \approx 3 \,\rm ms$. It could be a bit larger, 
$P_0 \leq 5 \,\rm ms$, for   
the canonical value of the pulsar braking index $n=3$ (the radio data 
then require $\tau_{\rm sd} \leq 20 \,\rm yr$).
In any case, the derived initial spin period  
of the Crab pulsar is much shorter than $P_0\approx 19\,\rm ms$ 
predicted previously (see Manchester \& Taylor 1977). Obviously, 
the reason for such puzzling discrepancy is to be understood prior to 
consider the model as realistic.

\begin{figure}[htbp]
\hspace{2cm}\resizebox{5cm}{!}{\includegraphics{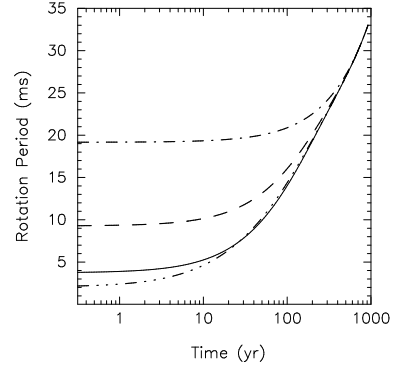}}
\caption
{The rotation period $P(t)$ of the pulsar resulting from
numerical integration of the equation ${\rm d}X /{\rm d}t = - K \,X^n$
for the normalized frequency $X\equiv \Omega/\Omega_\ast$, assuming $K(t)= K_0
[1+A\exp (-t/t_{\rm K})]$ with different values of parameters $A$ and
$t_{\rm K}$, and for different braking indices $n$:
 $A=0$ and $n=2.5$ (dot-dashed);
$A=4.1$, $t_{\rm K}= 170\,\rm yr$ and $n=2.25$ (dashed);
$A=6.2$, $t_{\rm K}= 155\,\rm yr$ and $n=2.25$ (solid).
The 3-dot--dashed line is calculated for $A=4.1$, $t_{\rm K}= 170\,\rm yr$,
but assuming a time-dependent $n(t)$ linearly decreasing from the initial
$n_0=2.6$ to $n_{\ast}=2.25$. In all cases the current value of the parameter
$\ddot{\Omega}_{\ast} \Omega_{\ast}/(\dot{\Omega}_\ast)^2=2.5$ is provided.
}
\end{figure}

\section{Discussion}

This puzzle is solved if we remember that    
$P_0$ has been previously derived from the relation  
\begin{equation}
t_{\ast}=\tau_{\ast} [ 1 -(\Omega_\ast/\Omega_0)^{n-1}\,]\; ,
\end{equation}
with $\tau_{\ast}=\Omega_\ast/(n-1)\dot{\Omega}_{\ast}$, which
is found after integration of the torque equation 
 $\dot{\Omega} = -K \, \Omega^n$ {\it assuming} that $n$ and 
$K$ do not depend on time. In that case the braking index $n$ 
can be expressed through 
the spin frequency $\Omega_{\ast}$ and its time 
derivatives at present as 
$n=\ddot{\Omega}_{\ast} \Omega_{\ast}/(\dot{\Omega}_\ast)^2$. For the 
Crab pulsar the measurements  result in $n = 2.5$ (Groth 1975), so  
for $t_\ast\sim 900\,\rm yr$  and $\tau_{\ast}= 1640\,\rm yr$ Eq.(6) 
predicts    
$P_0\approx 19 \,\rm ms$, which then implies also that  
$\tau_{\rm sd} \simeq 700\,\rm yr$.

Reliability of these predictions, however, is questionable
if we keep in mind that Eq.(6) is only an {\it approximate} 
formula for the pulsar age, and as such, it tells us that 
the model of a rotating magnetic dipole (Pacini 1967, Gunn \& Ostriker 1969)
does provide a good agreement with reality because 
$t_{\ast} \geq 0.6\tau_{\ast} $ for any $\Omega_0$.
However, attempts to treat this formula as a strict equation 
appropriate for {\it derivation} of $\Omega_0$ would overestimate 
its real accuracy,
 and can be largely misleading because it becomes practically  
independent  of $\Omega_{0}$  
already at times when $\Omega_{\ast} < \Omega_{0}/2$.
Actually, the {\it equation} (6) has to push  $\Omega_{0}$ close to the current 
$\Omega_{\ast}$ in order to be able to formally compensate the difference 
between $\tau_{\ast}$ and the known age $t_\ast$. 
Instead of that, a mathematically correct approach is to  
attribute this difference to the correction terms due in the 
approximate formula (6). 
For example, assuming that $n= \,\rm const$ but $K=K(t)$, we would derive  
$t_{\ast}=\tau_{\ast} [ 1 -(\Omega_\ast/\Omega_0)^{n-1}]\times C_{\ast}$ 
where $C_\ast=t_{\ast} K(t_\ast)/
\int_{0}^{t_\ast}K(t){\rm d}t$.
For $K(t)$ gradually declining in time, e.g. because of decreasing magnetic 
field $B_{\rm n}(t)$ of the neutron star ($K\propto B^2$), 
a correction term $C_\ast\geq 0.6 $ would readily allow 
any $\Omega_0\gg \Omega_\ast$. Some formal examples are shown in Fig.~2.   
Thus, the knowledge of $\Omega_\ast$, $\dot{\Omega}_\ast$ and 
$\ddot{\Omega}_{\ast}$ at present may not be sufficient for determination of 
$\Omega_0$ significantly larger than $\Omega_\ast$. Information about 
the initial 
parameters of pulsars can be found in the radio spectra of plerions  
which they create.

The results found for the Crab nebula allow
important conclusions about its history.
The initial rotation energy of the  
neutron star, and hence also of the explosion that  produced the star,
reach the level $\geq 10^{51}\,\rm erg$.  
This confirms the earlier suggestion of Chevalier (1977)   
that Crab Nebula was born in a normal Type II supernova event.  

Meanwhile, the total energy residing in the Crab nebula 
presently is  only $\simeq 4\times 10^{49}\,\rm erg$ 
(e.g. Davidson \& Fesen 1985). Where is   
then the initial $10^{51}\,\rm ergs$ of the pulsar?  The answer becomes 
clear if we take into account  that this energy has 
been injected into the nebula 
mostly by electrons ($\eta_{\rm e}\approx 1$, KC84) with  
a low-energy `cutoff' (!) and  
the mean energy $\overline{\gamma} \simeq \gamma_{\rm w}(t)$ that  at 
times $t\ll t_\ast$ was  
much higher than $\gamma_{\rm br}(t)$ down to which they 
have radiatively cooled on timescale $\Delta t\leq t$. 
 Because of that,   basically all of the injected energy  
has been radiated away, and only a tiny fraction 
$\varepsilon(t) \sim \gamma_{\rm br}/\gamma_{\rm w}\ll 1$ 
 of the pulsar spin-down power $L_{\rm sd}(t)$ was depositing in the nebula.  
The total energy accumulated  in the nebula by time $t$ would 
be then $W(t)\propto t^{2m} (1+t/\tau_{\rm sd})^{s-k}$.
 Remarkably,
because   $\gamma_{\ast \rm w}$ is still larger than $ \gamma_{\ast \rm br}$,   
this energy might be {\it increasing} 
even currently if $2m-k+s>0$. In particular,  in the case of 
$m=m_{\rm eq}\simeq 1.4$
predicted by the model for the  magnetic field 
evolving close to equipartition with the electrons, 
$2m-k+s>0$ if $s> -0.5$. Numerical calculations confirm this unexpected  result.
The increase of $W(t)$ at 
times $t\sim t_\ast \gg \tau_{\rm sd}$ (to be compared with
RC84) could explain the expansion  
of the nebula with a noticeable recent  (Trimble 1968), or perhaps even 
current, acceleration.

The principal suggestion of RG74 that the magnetic field energy  $W_{\rm B}$
in the Crab nebula is presently close to equipartition with 
$W_{\rm e}$ of the electrons
 not just accidentally, but because magnetic fields can effectively 
reenergize the electrons 
when $W_{\rm B}$ exceeds $ W_{\rm e}$, becomes important  for explanation of   
the observed small  decline rate   
of radio fluxes, $ \simeq (0.16 - 0.19)\,\%$ per year (Aller \& Reynolds 
1985, Vinyaikin \& Razin 1979).
For $m=m_{\rm eq}\simeq 1.4$, the calculated decline rate
\begin{equation} 
-{\rm d}\ln S_{\nu}/{\rm d}t  
\approx [m (1+\alr) + 2 a\alr(1+\delta )] /t_{\ast} \; .
\end{equation}
 is in a reasonable agreement with observations only if $a\ll 1$. 
Adiabatic reenergization of radio electrons in the regions of magnetic 
field compression could significantly compensate the 
expansion losses, effectively reducing the adiabatic parameter from   
$a=1$ relevant for a simple homogeneous model considered here, 
to much smaller values in the case of a spatially inhomogeneous study. 
Note that  the regions of magnetic field 
compression, like those around numerous dense optical   
filaments in the nebula, may be distinguished  by an enhanced 
radio intensity, e.g. as diffuse {\it radio} filaments (Swinbank 1980), 
but importantly, they will have practically the same  
$\alr\simeq 0.27$ as the rest of the nebula (Bietenholz et al. 1997), 
because the adiabatic energy gain (or loss) process 
does not change the power-law index of the electrons.

For the Crab pulsar, a speed of rotation 10 times higher than 
presently implies a very high initial spin-down luminosity  $L_{0} \sim 
10^{42}\,\rm erg/s$. The pulsar thus was injecting into
the young nebula an energy about $3\times 10^{49}\,\rm ergs$ per year, 
which was emitted mostly in the 
form of X-rays and $\gamma$-rays that could not be detected at that time. 
However, during the first years after the explosion, a significant part  
of this energy should be absorbed in the then opaque fast 
expanding shell of supernova ejecta between the nebula and an 
observer.  This additional energy input into the thermally 
emitting shell can be the reason why the light curve of the supernova 1054 
was declining much slower than that of a typical Type II event (Minkowski 1971).

The high initial luminosity of the pulsar could have also a strong
impact on the dynamics of the ejecta outside the nebula. 
Interesting consequences for the possible
progenitor of the Crab Nebula could be deduced from the high angular 
momentum of the pulsar, and hence of the progenitor star itself. 
Thus, the fast initial spin of the Crab pulsar as appears evident from 
the radio spectrum of the Crab nebula, suggests new approaches
for understanding the origin and history of this prototype plerion. 

\begin{acknowledgements}
I thank F. Aharonian, J. Arons, M. Bietenholz, W. Kundt,
R. Tuffs, and H. V\"olk for helpful discussions of different
aspects of this work, and R. Bandiera for very useful comments on 
the manuscript of the paper.
The work was supported by the  grant No. 05-2HD66A(7) of the German
BMBF.
\end{acknowledgements}

\end{document}